# Age-Associated Disorders As A Proxy Measure Of Biological Age:
## Findings From the NLTCS Data

A. Kulminski, A. Yashin, S. Ukraintseva, I. Akushevich, K. Arbeev,
K. Land and K. Manton.

(Center for Demographic Studies, Duke University)

**Abstract**

**Background:** The relative contribution of different aging-associated processes to the age phenotype may differ among individuals, creating variability in aging manifestations among age-peers. Capturing this variability can significantly advance understanding the aging and mortality. An index of age-associated health disorders (deficits), called a "frailty index" (FI), appears to be a promising characteristic of such processes. In this study we address the connections of the FI with age focusing on disabled individuals who might be at excessive risk of frailty.

**Methods:** The National Long Term Care Survey (NLTCS) assessed health and functioning of the U.S. elderly in 1982, 1984, 1989, 1994, and 1999. Detailed information for our sample was assessed from about 26,700 interviews. The individual FI is defined as a proportion of deficits for a given person. We perform cross-sectional empirical analysis of the FI age-patterns.

**Results:** FI in the NLTCS exhibits accelerated (quadratic) increase with age. Deficits might accumulate faster among the elderly who, at younger ages, had a low mean FI ("healthy" group) than a high FI ("disabled" group). Age-patterns for "healthy" and "disabled" groups converge at advanced ages. The rate of deficit accumulation is sex-sensitive. Convergence of the (sex-specific) FI for "healthy" and "disabled" groups in later ages determines biological age limits, associated with given levels of health-maintenance in the society, which correspond to 109.4 years for females and 92.5 years for males.

**Conclusions:** The FI can be employed as a measure of biological age and population heterogeneity for modeling aging processes and mortality in elderly individuals.

# 1. Introduction

Despite controversy in operational definition of frailty (1-6), current view on this phenomenon as on an organism's state with increased vulnerability to stressors and even decline or deregulation of physiologic functions (7) suggests various ways to assess frailty. One of them was suggested by Rockwood and Mitnitski and colleagues (1,8,9) who argued that health and quality-of-life deficits (i.e., signs, symptoms, impairments, etc.) accumulated by individuals during their life course could be considered as indicators of physiological frailty, biological aging (BA), and population heterogeneity.

In this paper, we apply the Rockwood-Mitnitski approach to constructing a cumulative index of deficits, called a frailty index (FI), using the National Long Term Care Survey (NLTCS). The NLTCS is a nationally-representative, longitudinal survey that assesses the health and functioning of U.S. elderly (65+) individuals over 18 years (1982, 1984, 1989, 1994, and 1999) with focus on disability (see (10) for details). To define a FI, we use the same, or similar, health deficits as assessed in the Canadian Study of Health and Aging (CSHA) (11,8). We will, thus, validate prior findings using a new population-based database and will focus on the connections of FI and age. Since disabled elderly are likely to be at excessive risk of frailty, the focus of the analysis on such category of individuals appears to be promising.

# 2. Methods

**The Frailty Index (FI).** The NLTCS contains a wide set of self-reported measures on health and functioning. Consistent with the view of the FI as a measure of functional complexity, Rockwood (9) argue that only the proportion of deficits constituting the FI is important in its relation to aging and mortality – not their specific substance. This provides flexibility in choosing deficits to construct the FI. Nevertheless, to be able to validate prior findings, we restrict ourselves to deficits similar to those assessed in the CSHA (8). Specifically, we selected 32 questions, presented in all waves, and grouped them according to missing rate: (i) difficulty with eating, dressing, walk around, getting in/out bed, getting bath, toileting, using telephone, going out, shopping, cooking, light house work, taking medicine, managing money; (ii) arthritis, Parkinson's disease, glaucoma, diabetes, stomach problem, history of heart attack, hypertension, history of stroke, flu, broken hip, broken bones; (iii) vision problem; self-rated health; and (iv) trouble with bladder or bowels, dementia, hearing problem, visit of hearing therapist, dentist, and foot doctor. Following Mitnitski-Rockwood's approach, we define the FI as an unweighted count of such deficits divided by the total number of all deficits considered for a person. For instance, if an individual has been administered 30 questions and responded positively (indicating that there is a deficit) to 5 and negatively (no deficit) to 24 then his/her FI is $5/29 \approx 0.172$.

**Missing Data.** Complete information was gathered in the NLTCS on questions covering disability, part of which is represented in first group (13 measures). The second group (11 measures) represents answers with very low percentage of missing data ranging from 0.07% to 1.3%. In the third group (2 measures), the variability of the proportion of missing answers across the five NLTCSs is slightly larger (0.6% to 3.7%). The fourth group (6 measures) represents questions with low proportions of missing data (about 0.5%), but which were not administered to all NLTCS participants. Since, for most questions, the proportion of missing data is small, the maximum number of available responses (i.e., from questions administered to NLTCS participants) is 30 for all waves. We constructed two FIs: one covering all 32 deficits and the other only the first three groups (i.e., 26 deficits).

## 3. Results

We first evaluate the FI age-patterns for each NLTCS wave. Despite the relatively large samples, estimates for single years of age are not sufficiently precise at the advanced ages (90+) where there are less than 100 cases per year. To improve statistical precision, and to smooth estimates, we used two-year age groupings in our analyses. Figure 1 shows the two-year age-patterns of the full (32 deficits) FI for five waves. The 26-deficit FI shows a similar age-pattern and thus is not depicted.

**Figure 1 is about here.**

Visual inspection of the age-patterns in Figure 1 reveals a nonlinear (accelerated) increase of the FI with age. Sex-differences of the 2-year FI age-patterns were not statically significant. Averaging the FI over 5 years of age (Table 1) shows that statistically significant differences between FIs for males and females are seen only for the 90-94 age group of the 1982 NLTCS and for 3 age groups (70-74, 80-84, 90-94) of the 1994 NLTCS. For the entire sample (65+) mean FIs for males and females are statistically different for each NLTCS, being lower for males than for females.

**Table 1 is about here.**

To find the best description of the age-patterns in Figure 1, we estimated several functions: linear, log-linear (or exponential), power, and quadratic. In all five NLTCS waves, and for FIs with both 32- and 26-deficits, the best fit is obtained by the quadratic function, $FI = U + B_1 \times Age + B_2 \times Age^2$, as determined by comparisons of coefficients of determination ($R^2$). T-test shows statistical significance for all coefficients except for $B_1$ and $U$ for the 1994 wave. Because the quadratic function has three parameters, the standard errors of its coefficients are larger than for the log-linear ($\ln(FI) = U + B_1 \times Age$) function (Table 2). For comparison, Table 2 also shows $R^2$ for linear functions in parenthesis. Two-year averaging significantly improved these estimates increasing the percentage of the total variation in dependent variables explained by nonlinear relations between age and FI by up to 50%. The use of five-year age categories did not noticeably improve fits. Thus, a quadratic function accurately describes the FI age-patterns in NLTCS data (Figure 1). The best fit was obtained for 1989 ($R^2 = 98\%$).

**Table 2 is about here.**

Despite the qualitative (shape) similarity of the age-patterns, there are quantitative differences among the waves. The largest mean age-specific FIs are for the 1982 NLTCS (Figure 1). Their difference from those of the other NLTCS waves is likely due to over-sampling of disabled individuals in the 1982 community questionnaire (91.2% in 1982 vs. 83.5% in 1984). Deviations of the 1982 NLTCS FI estimates from the exponential pattern of the age specific FI in the CSHA is also the largest – that is also reflected in the regression coefficients (Table 2).

In 1994, the NLTCS design was changed by adding the HS. The community questionnaire was completed by 1,303 persons (of 1,762 in the HS) in 1994 and by 1,196 (of 1,545 in the HS) in the 1999 wave. Since individuals in the HS were designated before the survey to receive a detailed interview, the proportion of non-disabled individuals in these groups is significantly lower than in the remaining ("disabled") group (DG) of individuals (selected for a community interview by the screener), being closer to the proportion in the U.S. elderly. Specifically, according to the age-adjusted estimates in (10), the prevalence of non-disabled elderly individuals in the national U.S. population in the 1999 was 80.3% and in 1994 was 77.5%. The prevalence of non-disabled respondents to the community questionnaire in the HS of the NLTCS without age standardization is 65.5% in 1999 and 80.5% in 1994. The over-sampling of "healthy" individuals reduces the mean FI for the 1994 and 1999 waves – especially at "younger" ages providing good agreement with results from CSHA.

The presence of the HS in the 1994 and 1999 waves provides an opportunity to estimate the difference between survey and community samples distinguishing the age-patterns of the DG and HS. Figure 2 shows that the age-pattern for the DG shifts up becoming closer to the 1982-1989 patterns. Meanwhile, age-patterns for the HS shift down exhibiting smaller mean FIs than those assessed from CSHA for all age groups. Again, better fits are obtained for the quadratic function except for the 1999 HS, for which the log-linear (exponential) fit is better (Table 3).

**Figure 2 is about here.**
**Table 3 is about here.**

Figure 2 suggests that individuals in the HS (small FI at young ages) accumulate deficits faster than those in the DG (large FI). To increase statistical power, we pooled data for 1994 and 1999 waves and averaged the FI over 5 years of age. Figure 3 exhibits the 1994&1999 FI age-patterns for the entire sample (left panel) and for both sexes (right panel) along with their nonlinear fits (Table 4). Figure 3 clearly shows that individuals from the HS accumulate deficits faster than those from DG. The rate of deficit accumulation varies by sex. This is also seen considering each wave separately and averaging FI over larger age intervals (Table 5). Specifically, males in the HS have smaller FI at younger ages than females. However, males accumulate deficits faster than females resulting in convergence of their FI age-patterns and crossing at advanced ages (~85).

**Figure 3 is about here.**
**Table 4 is about here.**
**Table 5 is about here.**

### 4. Discussion and Conclusions

Our analyses show that the mean FI increases with age, and that this increase is nonlinear (with acceleration), i.e., older people accumulate more deficits per year than younger. In most cases, the age-pattern is best described as a quadratic function. This means that the rate of increase also increases with age (in a linear fashion) stressing the nonlinear nature of deficit accumulation. The best fits, when quadratic fits are insignificant, were exponential. Correlation of the FI with age and similarity between the FI and mortality age patterns suggest that the FI could be used as an adequate indicator of BA (11). Although, usually, BA indicators are expected to have linear relation with chronological age (12), it can be argued that the relation should be nonlinear. One argument for that is the high plasticity and age-dependence of mortality rate *variation* in experiments with anti-aging interventions aimed to increase longevity (13). Valid biomarkers of aging must capture these properties, i.e., they must have a nonlinear relation with chronological age. Another argument is that the overall rate of somatic aging might be the product of a combination of rates of different biological processes with distinct age dynamics which can result in nonlinear change of BA indices with chronological age (14-17).

The FI appears to have the potential to differentiate aging processes at individual level. Consequently, FI becomes useful characteristic describing population heterogeneity in various models of aging and mortality, which can be implemented using, for instance, microsimulation procedures designed to assess the impact of individual states (18).

Our results reveal large differences between the FI age-patterns for the 1982, 1984, and 1989 NLTCS waves as compared to the 1994 and 1999 waves which appear due to the presence of a "healthy" sample in the two later waves. Only the patterns for the last two waves resemble those from the CSHA. The CSHA sample is representative of elderly (65+) Canadians who are being screened according to cognitive function (11).

It was argued (8) that the FI age-patterns are largely independent of survey design. Our results show that survey design is a serious issue in constructing FIs even using similar community-based samples. This occurs because intentional, or unintentional, screening can result in over-representation of individuals with certain health/quality-of-life deficits. The NLTCS community sample is an example of an intentional selection of disabled individuals by screening and sample selection procedures. We dealt with that feature of the NLTCS sampling by stratifying on the HS versus the non-HS (DG). The CSHA focuses on selection of cognitively impaired elderly which, as a consequence of their mental abilities, have larger proportions of health deficits and poorer quality of life measures than those with intact cognitive functions (19). Therefore, even if a survey does not directly focus on specific aspects of the individuals' health which constitute large part of the deficits included in the FI definition, such individuals can be over-sampled in the survey (i.e., the survey sample can approximate a non-community setting) thereby increasing the mean FI. Consequently, it is reasonable to expect that mean FIs for survey participants can be larger than for community-dwelling individuals provided that such deficits are part of the FI definition.

The presence of the HS in the NLTCS allowed us to directly confirm this fact. Individuals for the DG were selected following standard NLTCS procedures (i.e., screening in disabled individuals), while for the HS they were selected irrespective of disability. Since the screener NLTCS participants were primarily selected from the U.S. Medicare eligible persons to provide nationally representative sample according to demographic factors, the likelihood of systematic bias resulting in selection of individuals for HS with specific health problems is low.[1] Our analysis shows that the FIs for the general population of community-dwelling elderly should be lower than those estimated using particular surveys.

NLTCS data provide evidence on complex (nonlinear) relationships between the FI, sex, and age. To understand this complexity, we make four observations. First, the mean FI for males is smaller than for females for each NLTCS wave. This agrees with other findings (8). However, this difference is not large. Second, there is not, generally, a statistically significant sex difference between age-specific FIs. Third, there is no overall tendency that the FI for males is less than for females. Fourth, analysis of the sex-specific FIs for different age groups shows two opposite tendencies in the sex-sensitivity of the FI behaviors with age (Table 1). Specifically, at younger ages in the early waves, FIs are nearly identical but have tendency to diverge with age.

---

[1] This fact has been also verified by comparing the FI age-patterns for the HS and for the U.S. community-dwelling elderly. The latter sample was obtained from respective NLTCS wave (1994 or 1999) using weights developed by the U.S. Bureau of the Census and the Center for Demographic Studies (Duke University) to produce national estimates. Both (weighted and HS) estimates show excellent agreement, especially at younger ages.

For the two latest waves, there is a tendency towards convergence of these indices at the extreme ages. Since the two later waves have a smaller proportion of disabled individuals due to the presence of the HS, it is reasonable to assume that the latter fact is responsible for such a change. Indeed, when considering the DG and HS separately (Figure 3), the qualitative change of the FI with age becomes more pronounced. Males and females in the DG have essentially similar FIs at younger ages — the opposite fact is seen for the HS. This is a clear nonlinear effect when the relation between FIs for males and females is FI- and age- dependent. A consequence is that in different settings (e.g., institutional vs. hospital vs. community) the relation between FI for males and females can be qualitatively different.

The intriguing finding of our study is that FIs for HS and DG converge at the oldest-old ages, i.e., the rate of deficit accumulation for individuals in the HS is larger than in the DG. This finding suggests that aging process itself rather than particular pathologies plays pivotal role in the risk of death at extreme ages. Such behavior becomes even more pronounced in male and female sub-groups. The rate of deficit accumulation for females is larger than for males for the DG. For the HS, we see the opposite situation. As a consequence, the difference in the rates results in divergence of FI age-patterns for males and females in DG and in their convergence in the oldest-old ages for the HS. In other words, for large FI at younger ages the FI age-patterns appear flatter than those for small FI at younger ages. Figure 3 (right panel) also suggests that sensitivity to the quantity of the accumulated deficits is higher for males than for females. This follows from the fact that males and females accumulate deficits with age at different rates and differently in the DG and the HS. Changes in rates between DG and HS are larger for males than for females.

These findings provide further support for considering the FI as a measure of BA. Since humans have limited life spans (i.e., no individuals live an unlimited time, although, the life-span-limits might change with improvements in economic standard of living, social conditions, and medical progress (20)), the FI – as a BA indicator – should be able to characterize BA limits associated with given level of health-maintenance in the society (21). Specifically, in a community setting (approximated by the HS in our analysis), males and females have smaller mean FIs, especially at younger ages, than age-peers in non-community groups (e.g., the DG). However, the FI increases with age faster in the HS than in the DG. This may be due to the presence of a BA limit. Our data provide an opportunity to determine a BA limit from the extrapolated fits. For the HS and DG samples, this occurs at age 104.5 years at $FI = 0.435$. Individuals with both elevated (DG) and normal (HS) FI level at younger ages can reach this BA limit. However, individuals from the HS would have to age faster to reach the same limit.

Our data suggest that males and females have different BA limits. Interpolation of the female-specific fits for the HS and DG to extreme ages provides a reasonable estimate for the females' BA limit of approximately 109.4 years ($FI = 0.456$). For males, we obtain a lower limit of 92.5 years. This estimate, however, was imprecise due to the small sample of males at those ages. The difference in the BA age limits for males and females may be the reason why there are opposite tendencies in the sex-specific FI age-patterns in the DG and HS. Indeed, since males have a lower BA limit, those who are in the HS accumulate deficits with age faster than females. For the same reason, males in the DG accumulate deficits with age slower than females.

The presence of BA limit does not mean that longevity cannot be extended beyond certain age. It rather exhibits systemic feature of the aging process and indicates the need of development of adequate systemic methods of copying with this phenomena. Such methods focusing on slowing down the rate of deficit accumulation will result in extension of both life span and active live life span, even if the BA limit will remain unchanged. Consequently, health-care providers should focus their efforts not only on individuals with serious health problems, but also on "healthy" individuals (i.e., with mild health problems) at younger ages to reduce the likelihood of fast nonlinear accumulation of heath deficits at advanced ages. At the same time the progress in medical technology may affect the BA limit as well. How all such transformations will affect the quality of life at late ages deserves separate study.

Acknowledgments.

The research reported here was supported in part by P01-AG-017937-05 grant from the National Institute on Aging (NIA). A.K. also acknowledges support from K12-AG-000982-05 NIA grant.

**Table 1. Sex-specific FIs and the number of males (M) and females (F) for each NLTCS. An asterisk (plus) denotes $0.01 < p \leq 0.05$ ($p \leq 0.01$). Other sex differences are insignificant.**

| NLTCS | Sex | N | 65+ | Age groups | | | | | | |
|---|---|---|---|---|---|---|---|---|---|---|
| | | | | 65-69 | 70-74 | 75-79 | 80-84 | 85-89 | 90-94 | 95+ |
| 1982 | M | 2166 | .266 (.003)$^+$ | .244 (.007) | .262 (.006) | .259 (.007) | .289 (.009) | .289 (.010) | .311 (.018)* | .337 (.027) |
| | F | 3921 | .277 (.002)$^+$ | .243 (.005) | .252 (.005) | .274 (.005) | .282 (.005) | .311 (.006) | .352 (.010)* | .374 (.021) |
| 1984 | M | 2038 | .250 (.003)* | .229 (.007) | .238 (.007) | .244 (.007) | .264 (.008) | .283 (.011) | .321 (.018) | .355 (.034) |
| | F | 3891 | .259 (.002)* | .226 (.005) | .231 (.005) | .244 (.005) | .269 (.005) | .297 (.006) | .349 (.010) | .344 (.018) |
| 1989 | M | 1470 | .241 (.004)$^+$ | .235 (.010) | .214 (.007) | .234 (.008) | .259 (.010) | .274 (.013) | .294 (.021) | .341 (.055) |
| | F | 2992 | .258 (.003)$^+$ | .230 (.007) | .226 (.006) | .247 (.005) | .261 (.006) | .283 (.007) | .337 (.010) | .376 (.020) |
| 1994 | M | 1736 | .191 (.004)$^+$ | .150 (.008) | .163 (.008)* | .187 (.006) | .206 (.009)$^+$ | .255 (.013) | .255 (.023)* | .335 (.031) |
| | F | 3336 | .221 (.003)$^+$ | .159 (.007) | .184 (.006)* | .195 (.004) | .238 (.006)$^+$ | .272 (.007) | .317 (.012)* | .339 (.012) |
| 1999 | M | 1805 | .196 (.004)$^+$ | .162 (.008) | .148 (.008) | .190 (.008) | .212 (.007) | .247 (.011) | .256 (.020) | .353 (.023) |
| | F | 3341 | .220 (.003)$^+$ | .167 (.007) | .165 (.006) | .202 (.005) | .224 (.005) | .266 (.008) | .290 (.012) | .344 (.013) |

**Table 2. Coefficients for the log-linear (Ln) and quadratic (Q) functions along with coefficients of determination ($R^2$) for each NLTCS wave. $R^2$ in parentheses is given for linear functions for the sake of comparison. Estimated coefficients are significant at the 0.05 level or better. Superscript "#" denotes insignificant estimates.**

| NLTCS | Fit | $B_1$ (SE)$\times 10^2$ | $B_2$ (SE)$\times 10^4$ | $U$ (SE) | $R^2$, % |
|---|---|---|---|---|---|
| 1982 | Ln | 1.37 (.071) | | -2.346 (.058) | 96.1 (93.6) |
|      | Q  | -1.10 (.313) | 0.93 (.193) | 0.563 (.125) | 97.7 |
| 1984 | Ln | 1.63 (.139) | | -2.615 (.113) | 90.2 (87.7) |
|      | Q  | -1.90 (.597) | 1.46 (.368) | 0.846 (.239) | 94.2 |
| 1989 | Ln | 1.66 (.157) | | -2.655 (.128) | 88.2 (86.0) |
|      | Q  | -2.81 (.360) | 2.03 (.222) | 1.202 (.144) | 98.0 |
| 1994 | Ln | 2.72 (.116) | | -3.689 (.095) | 97.3 (96.3) |
|      | Q  | -0.58 (.517)# | 0.75 (.318) | 0.210 (.207)# | 97.4 |
| 1999 | Ln | 2.67 (.150) | | -3.667 (.122) | 95.5 (93.8) |
|      | Q  | -1.63 (.514) | 1.39 (.317) | 0.627 (.206) | 97.4 |

**Table 3. Coefficients for the quadratic and log-linear (denoted by "#") functions fitting data in Figure 2 for the HS and DG in 1994 and 1999 NLTCS waves. $R^2$ is also given for linear (Lin) function. For all estimates $p \leq 0.05$.**

| NLTCS | Group | $B_1$ (SE)$\times 10^2$ | $B_2$ (SE)$\times 10^4$ | $U$ (SE) | $R^2$, % Q | Ln | Lin |
|---|---|---|---|---|---|---|---|
| 1994 | HS | -3.66 (.95) | 2.73 (.61) | 1.30 (.37) | 95.1 | 92.3 | 86.2 |
|      | DG | -1.54 (.55) | 1.27 (.34) | 0.65 (.22) | 95.8 | 92.9 | 91.6 |
| 1999 | HS# | 4.83 (.46) | | -5.79 (.36) | 89.0 | 90.2 | 87.3 |
|      | DG  | -1.54 (.45) | 1.23 (.28) | 0.69 (.18) | 96.2 | 92.5 | 91.0 |

**Table 4. Coefficients of the best statistically significant (p<0.05) fits corresponding to the curves in Figure 3.**

| Group | Sex | Fit | $B_1$ (SE)$\times 10^2$ | $B_2$ (SE)$\times 10^4$ | $U$ (SE) | $R^2$, % |
|---|---|---|---|---|---|---|
| HS | M&F | Ln | 4.42 (.27) | | -5.45 (.22) | 98.1 |
|    | M   | Q  | -5.05 (.96) | 3.84 (.62) | 1.73 (.37) | 99.3 |
|    | F   | Ln | 4.02 (.31) | | -5.08 (.25) | 97.1 |
| DG | M&F | Q  | -1.51 (.23) | 1.24 (.15) | 0.66 (.09) | 99.7 |
|    | M   | Ln | 1.57 (.34) | | -2.68 (.27) | 81.0 |
|    | F   | Q  | -1.07 (.37) | 0.98 (.23) | 0.48 (.14) | 99.4 |

**Table 5. Mean FIs for males and females for age-specific age groups for HS and DG of 1994 and 1999 NLTCS waves.**

| Sex | Group | Age | 1994 | | 1999 | |
|---|---|---|---|---|---|---|
| | | | Mean FI | SE | Mean FI | SE |
| Male | HS | 65-74 | .065 | .005 | .084 | .007 |
| | | 75-84 | .097 | .007 | .129 | .007 |
| | | 85+ | .237 | .041 | .217 | .026 |
| | DG | 65-74 | .218 | .007 | .189 | .007 |
| | | 75-84 | .227 | .006 | .229 | .006 |
| | | 85+ | .270 | .011 | .267 | .010 |
| Female | HS | 65-74 | .093 | .006 | .107 | .007 |
| | | 75-84 | .125 | .006 | .138 | .006 |
| | | 85-94 | .187 | .019 | .216 | .016 |
| | | 95+ | .222 | .074 | .385 | .054 |
| | DG | 65-74 | .217 | .006 | .192 | .006 |
| | | 75-84 | .239 | .004 | .236 | .004 |
| | | 85-94 | .299 | .007 | .284 | .007 |
| | | 95+ | .343 | .012 | .343 | .013 |

**Figure legends.**

Figure 1. The two-year frailty index age-patterns for each National Long Term Care Survey along with model estimate of the frailty index age distribution for the Canadian Study of Health and Aging (CSHA; thick line *FI=exp(0.029Age-4.05)*, (8)). The standard errors ($\pm SE$) of means are shown by bars for the 1982 and 1999 waves.

Figure 2. The two-year frailty index age-patterns for the healthy sample (HS) and disability group (DG) for the 1994 and 1999 National Long Term Care Survey waves. The 95% confidence intervals (CI) of means are shown by bars for 1994 HS and 1994 DG.

Figure 3. The five-year frailty index age-patterns for the healthy sample (HS) and disability group (DG) for pooled 1994&1999 data for entire sample (left panel) and for both sexes (right panel). Dashed-dotted line denotes extrapolation of the respective fitted curves. Bars show 95% CI. Dashed (continuous) line on the right panel denotes fits for males (females).

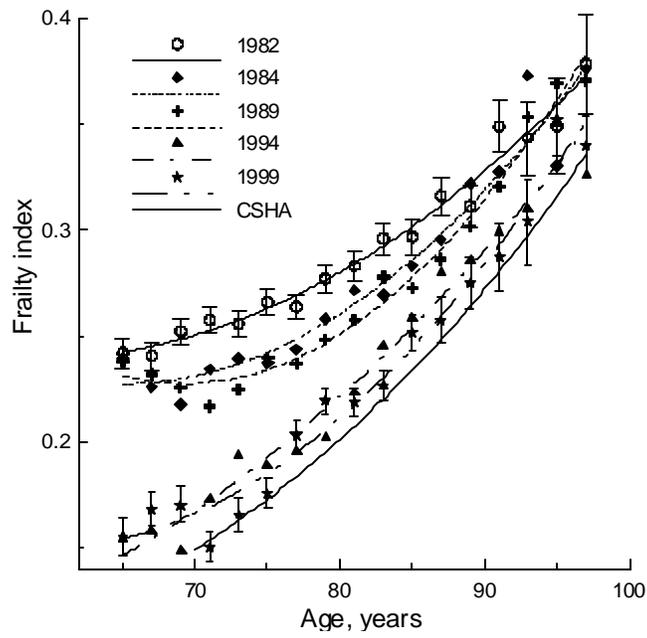

**Figure 1.**

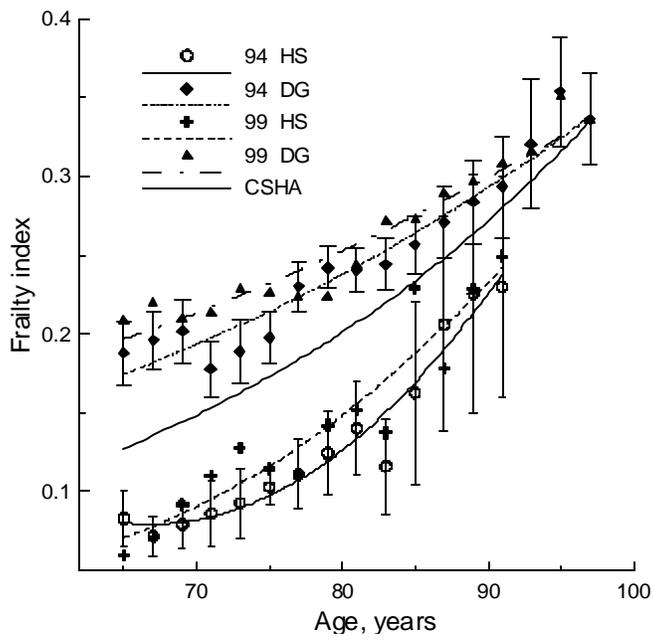

**Figure 2.**

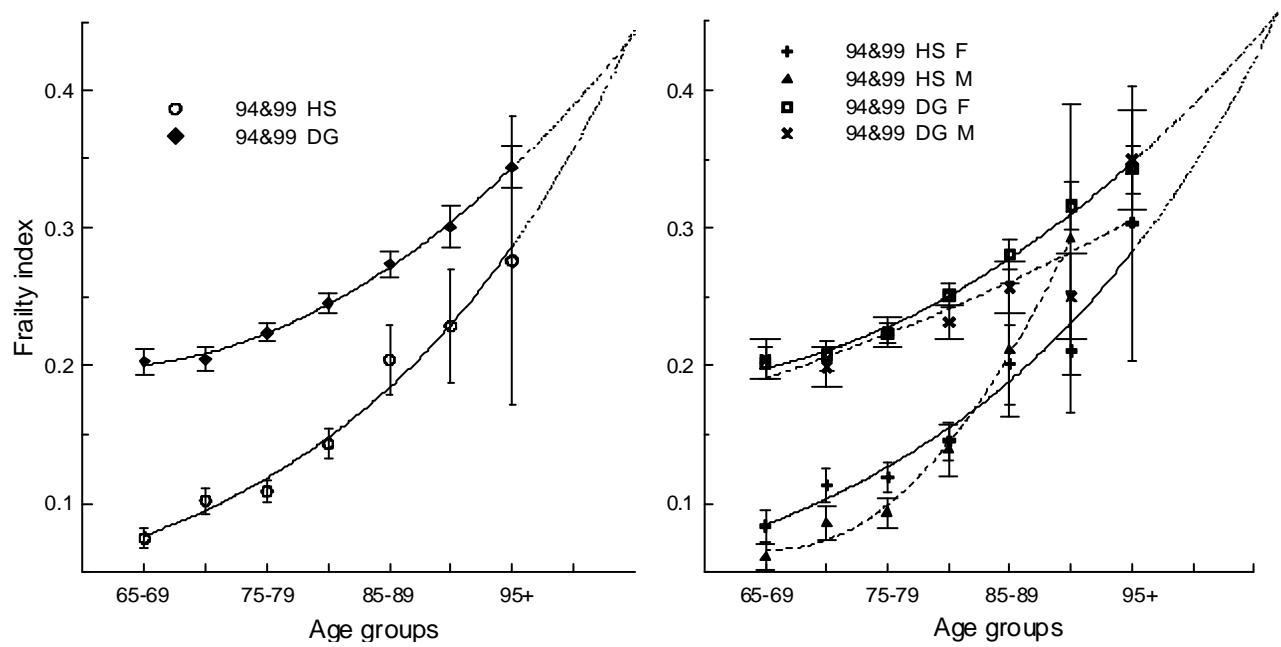

**Figure 3.**